\documentclass[twocolumn,amsmath,amssymb,prl,10pt,nofootinbib,superscriptaddress]{revtex4}
\usepackage{ graphicx, float,amsmath, amsmath, amssymb}
\usepackage[export]{adjustbox}
\usepackage[breaklinks, colorlinks, citecolor=blue]{hyperref}
\usepackage[labelsep=period]{caption}
\usepackage[normalem]{ulem}
\usepackage{mathtools}
\usepackage{siunitx}
\usepackage{soul}
\usepackage{physics}
\usepackage{minitoc}

\usepackage{bm}
\usepackage{xcolor}

\definecolor{rosso}{cmyk}{0,1,1,0.4}
\definecolor{rossos}{cmyk}{0,1,1,0.55}
\definecolor{rossoc}{cmyk}{0,1,1,0.2}
\definecolor{blu}{cmyk}{1,1,0,0.3}
\definecolor{blus}{cmyk}{1,1,0,0.6}
\definecolor{bluc}{cmyk}{1,1,0,0.1}
\definecolor{verde}{cmyk}{0.92,0,0.59,0.25}
\definecolor{verdec}{cmyk}{0.92,0,0.59,0.15}
\definecolor{verdes}{cmyk}{0.92,0,0.59,0.4}
\definecolor{Gray}{gray}{0.95}

\font\tenrsfs=rsfs10 at 12pt
\font\sevenrsfs=rsfs7
\font\fiversfs=rsfs5
\newfam\rsfsfam
\textfont\rsfsfam=\tenrsfs
\scriptfont\rsfsfam=\sevenrsfs
\scriptscriptfont\rsfsfam=\fiversfs

\hypersetup{
     colorlinks = true,
     linktocpage =true,
     linkcolor = verdes,
     citecolor = verdes,
     anchorcolor = verdes,
     filecolor = blue,
     urlcolor = verdes,
     bookmarks = true,
     bookmarksopen = true,
     bookmarksnumbered = true,
     pdfstartview=FitH
     }

\newcommand{\bea}{\begin{eqnarray}}
\newcommand{\eea}{\end{eqnarray}}
\newcommand{\beq}{\begin{equation}}
\newcommand{\eeq}{\end{equation}}

\numberwithin{equation}{section}

\begin{document}
\title{The Grenoble Axion Haloscope platform (GrAHal): development plan and first results}

\author{Thierry Grenet}%
\affiliation{%
Institut N\'eel, Universit\'{e} Grenoble-Alpes, CNRS, Grenoble INP,\\38000 Grenoble, France
}

\author{Rafik Ballou}%
\affiliation{%
Institut N\'eel, Universit\'{e} Grenoble-Alpes, CNRS, Grenoble INP,\\38000 Grenoble, France
}

\author{Quentin Basto}%
\affiliation{%
Institut N\'eel, Universit\'{e} Grenoble-Alpes, CNRS, Grenoble INP,\\38000 Grenoble, France
}

\author{Killian Martineau}%
\affiliation{%
Laboratoire de Physique Subatomique et de Cosmologie (LPSC), Universit\'e Grenoble-Alpes, CNRS/IN2P3\\
53, avenue des Martyrs, 38000 Grenoble, France
}

\author{Pierre Perrier}%
\affiliation{%
Institut N\'eel, Universit\'{e} Grenoble-Alpes, CNRS, Grenoble INP,\\38000 Grenoble, France
}

\author{Pierre Pugnat}%
\affiliation{%
Laboratoire National des Champs Magn\'etiques Intenses (LNCMI),\\
European Magnetic Field Laboratory (EMFL), Universit\'e Grenoble-Alpes, CNRS, 38000 Grenoble, France
}

\author{Jérémie Quevillon}%
\affiliation{%
Laboratoire de Physique Subatomique et de Cosmologie (LPSC), Universit\'e Grenoble-Alpes, CNRS/IN2P3\\
53, avenue des Martyrs, 38000 Grenoble, France
}

\author{Nicolas Roch}%
\affiliation{%
Institut N\'eel, Universit\'{e} Grenoble-Alpes, CNRS, Grenoble INP,\\38000 Grenoble, France
}

\author{Christopher Smith}%
\affiliation{%
Laboratoire de Physique Subatomique et de Cosmologie (LPSC), Universit\'e Grenoble-Alpes, CNRS/IN2P3\\
53, avenue des Martyrs, 38000 Grenoble, France
}

\date{\today}
\begin{abstract} 
In this note we report on the development plans and first results of the Grenoble Axion Haloscope (GrAHal) project. It is aimed at developing a haloscope platform dedicated to the search for axion dark matter particles. We discuss its general framework and the plans to reach the sensitivity required to probe well known invisible axion models, over particularly relevant axion masses and coupling regions. We also present our first haloscope prototype and the result of its test run at liquid He temperature, setting a new exclusion limit $g_{a \gamma \gamma} \leq 2.2 \times 10^{-13}~ \text{GeV}^{-1}$ ($g_{a \gamma \gamma} \leq 22 \times g_{\text{KSVZ}}$) around 6.375 GHz ($m_a \simeq 26.37$ $\mu \text{eV}$).

\end{abstract}
\maketitle

\section{Introduction}

New particles in unexplored territories of mass and interactions are predicted in many extensions of the Standard Model (SM) of particle physics, in particular in the low energy frontier of weakly interacting sub-eV particles. These include the QCD axion and the axion-like particles (ALPs). An utmost interest of these particles is to provide serious candidates for the observed dark matter in the universe for specific ranges of mass and density.
The QCD axion is a pseudo-scalar particle associated with the spontaneous breaking, at an energy scale $f_a$, of a new chiral symmetry (the so-called Peccei-Quinn symmetry) initially postulated to explain the lack of experimentaly known charge-parity (CP) violation in quantum chromodynamics (QCD) \cite{Peccei:1977hh,Peccei:1977ur,Weinberg:1977ma,Wilczek:1977pj}. The aforementioned symmetry being anomalous, the axion couples to gluons and acquires a mass $m_a$ that scale with $f_{a}^{-1}$ at the QCD phase transition. 
Depending on the exact model, axions can interact with other particles, as the SM fermions and the photons with the typical scaling factor $f_{a}^{-1}$.
It was later realized that the QCD axion could be a serious Dark Matter (DM) candidate through coherent oscillations \cite{Preskill:1982cy,Abbott:1982af,Dine:1982ah} of the associated field. ALPs can also be produced from a misalignment mechanism and thereby provide contributions to DM.

A number of experiments have been proposed and conducted to search for the QCD axion and ALPs based on their (possible) interactions with the SM particles \cite{Sikivie2021}. Among those exploiting their coupling to two photons, 
the largest experimental effort is presently made towards the use of radio-frequency (RF) cavities for resonant conversion of axions into photons in the $10^{-6}-10^{-4}$~eV mass range, where QCD axions are supposed to contribute dominantly to DM. In the world several projects of this kind are underway aiming at different masses regions, namely: ADMX \cite{ADMX:2018gho, PhysRevD.103.032002}, HAYSTACK \cite{Brubaker:2016ktl, HAYSTACK2021}, CAPP \cite{Choi:2020wyr}, ORGAN \cite{McAllister:2017lkb}, CAST-RADES \cite{melcon2021results} and
QUAX$_{a\gamma}$ \cite{Alesini:2019ajt}, as well as MADMAX, a project of dielectric haloscope \cite{PhysRevLett.118.091801}.

Axions detection via a RF cavity haloscope relies on the assumption that they constitute a significant part of the DM galactic halo. In this case it is possible to detect their conversion into photons in a resonant cavity immersed in an intense magnetic field. The signal is power amplified as
\begin{equation}
P = g^2_{a \gamma \gamma} \rho_a \frac{1}{m_a} \frac{\beta}{\left(1 + \beta \right)^2} Q_0 B^2 VC \frac{1}{1+\left(\frac{2 (\nu - \nu_c)Q_0}{\nu_c (1+\beta)}\right)^2}
\label{eq:power}
\end{equation}
where $g_{a\gamma\gamma}$ is the axion-diphoton coupling constant, $\rho_a$ is the halo DM density, $m_a$ the axion mass, $\beta$ the cavity coupling to the receiver chain, $Q_0$ the unloaded cavity quality factor, $B$ the magnetic field, $V$ the cavity volume, $C$ the coupling factor of the electromagnetic mode, $\nu_c$ the cavity mode frequency and $\nu$ the measured frequency.

The instrumental challenge lies in the weakness of the expected signal (of the order of $10^{-24}-10^{-23}$~W) relative to the thermal radiation of the cavity. The signal-to-noise ratio (SNR) can be approximated by the Dick radiometer formula
\begin{equation}
SNR=\frac{P}{k_{B} T_{\text{syst}}} \sqrt{\frac{\Delta t}{\Delta \nu_a}}
\label{eq:snr}
\end{equation}
where $k_B$ is the Boltzmann constant, $T_{\text{syst}}=T+T_N$, i.e. the sum of the detector temperature and the amplifier noise temperature $T_N$, $\Delta t$ the integration time and $\Delta \nu_a$ the analysis bandwidth (set to the axion bandwidth). The SNR can be optimised by lowering $T_{\text{syst}}$, which involves cooling the cavity at ultra-low temperature and using amplifiers operating at the quantum limit of noise or even below.

The most sensitive results have been obtained so far by the ADMX experiment \cite{PhysRevD.103.032002}. Using quantum amplifiers, this experiment has shown that the technology is now mature and sensitive enough to test the so called axion DFSZ model~\cite{Dine:1981rt,Zhitnitsky:1980tq} through photon couplings in the $\mu \text{eV}$ range.

The Grenoble Axion Haloscope project (GrAHal) aims at developing a haloscope platform in Grenoble (France), able to run detectors of different sizes and designs for the search of galactic axions and ALPs at the best sensitivity in the $0.3 - 30$ GHz frequency range (1.25 - 125 $\mu \text{eV}$). It draws on Grenoble's internationally recognised know-how on key experimental aspects of the project: high magnetic fields and fluxes, ultra-low temperatures and ultra-low noise quantum detectors.



\section{The GrAHal project}


\subsection{Magnet systems}
 
\begin{figure}
    \centering
    \includegraphics[width=0.45\textwidth]{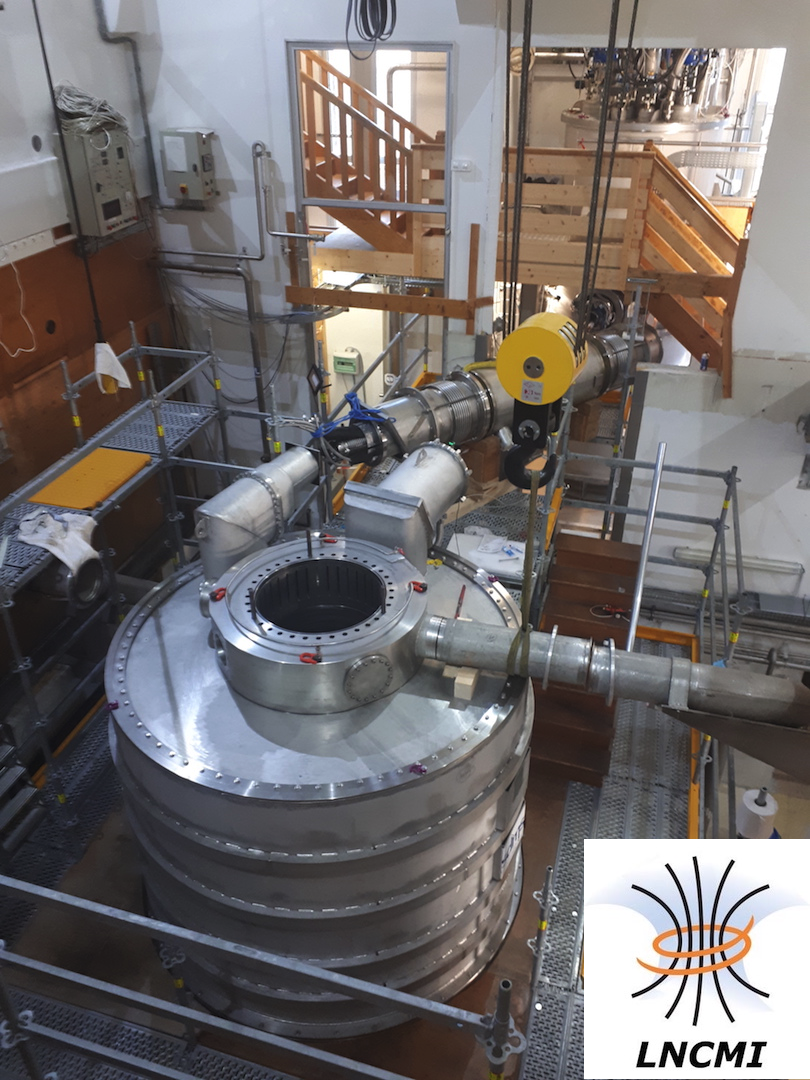}
    \begin{flushleft}
     \caption{Status of the construction of the Grenoble hybrid magnet. \underline{In the front:} assembled magnet cryostat with its warm bore diameter of 812 mm. It contains the superconducting coil immersed inside its 1.8 K He vessel and surrounded by three thermal shields at 4 K, 30 K and 100 K. On the top of the magnet cryostat, one of the water cooling pipe for resistive magnets is being connected. \underline{In the back:} the cryogenic line to be connected to the magnet cryostat and to the cryogenic satellite producing the superfluid He can also be seen. \underline{Final assembly:} beginning of 2022. \underline{First operation:} beginning of 2023.
     }
     \label{figure:hybrid_magnet}
     \end{flushleft}
\end{figure}

One of the key tools of GrAHal is the Grenoble hybrid magnet (cf. Fig.\ref{figure:hybrid_magnet}) close to completion at LNCMI \cite{LNCMIReport2020} and planned to be fully operational in 2023 after first cooling down to 1.8 K and commissioning tests in 2022. It is a modular experimental platform offering various possibilities of maximum field values and useful warm bore diameters.
From the combination of resistive polyhelix and Bitter coils inserted within a large bore superconducting one, a maximum field of at least 43 T will be produced in a 34 mm diameter aperture with 24 MW of electrical power. By combining the Bitter insert alone with the superconducting coil, another hybrid magnet configuration will allow to produce 17.5 T in a 375 mm diameter aperture with 12 MW. With the superconducting coil alone, a maximum field of 9.5 T in 812 mm diameter bore can be reached. Compared to other running haloscopes 
the superconducting outsert magnet alone offers an unprecedented value for the figure of merit B$^2$V $\approx 60$~T$^2$ m$^3$. Other configurations of magnetic field and warm bore diameter offered are listed in Table.\ref{table:Bvsdia_configs}. To highlight the unique opportunities offered by this magnet to probe a large domain of axion mass and diphoton coupling, the inner diameters of cylindrical RF-cavities that can be inserted inside the various magnet apertures are also given with their TM$_{010}$ resonance frequency and corresponding axion mass.\\

\begin{table}[htp]
\begin{center}\begin{tabular}{|c|c|c|c|c|}\hline Magnetic& Warm & RF-Cavity& Resonant & Axion \\ Field & Diameter & Diameter & Frequency & Mass \\ (Tesla) & (mm) & (mm)* & (GHz) & ($\mu eV$) \\\hline 43 & 34 & 8 & 29 & 118 \\ 40 & 50 & 23 & 10 & 41 \\ 27 & 170 & 110 & 2 & 8.6 \\ 17.5 & 375 & 315 & 0.7 & 3 \\ 9.5 & 812 & 675 & 0.34 & 1.4 \\\hline\end{tabular} 
\\ ~ \\ \textit{* From 1st cut design and after subtraction of the cryostat plus cavity walls thickness.}
\end{center}
\caption{Main characteristics of the modular Grenoble hybrid magnet user platform. For reference, corresponding inner diameters of cylindrical RF cavities are also given. Dedicated tuning mechanisms will allow frequency scans. The possibility to insert several matched cavities of smaller diameter for each hybrid magnet configuration is also considered.}
\label{table:Bvsdia_configs}
\end{table}

In addition to the Grenoble hybrid magnet platform, several superconducting coils at CNRS-Grenoble producing magnetic fields up to 20 T in 50 mm diameter 
can be used as test beds for RF cavities and allow the operation of several haloscopes in parallel. Finally, within four to five years, GrAHal will also benefit from the recently funded FASUM project  
offering a full superconducting coil producing 40 T in 30 mm aperture with the possibility of also using the 20 T outsert of 150 mm diameter bore alone.

\subsection{Detector developments}

Large magnet bores allow large cavity volumes to enhance the axion signal to be detected. These cavities can be developped in collaboration with IBS-CAPP in Korea \cite{Pugnat:395493}. The TM$_{010}$ frequency of a cylindrical cavity, and thus the axion mass searched for, scales inversely with the radius (cf. Table \ref{table:Bvsdia_configs}) and a cavity mode frequency can only be tuned over typically one octave. Thus, higher mode frequencies imply smaller volumes and degraded sensitivities. Probing the whole dark matter axion mass range will imply the development of more involved designs to circumvent this difficulty, like multiple-cell cavities \cite{2020_multiple-cell}, networks of single cavities, or detectors based on the conversion of axions to specific excitations such as e.g. plasmons \cite{PhysRevLett.123.141802}.

To increase sensitivity, RF cavities with quality factors of order $Q \sim 10^5-10^6$ (higher than the values obtained with copper) will have to be built, which will require dedicated developments such as superconducting cavity walls compatible with the magnetic field environment.

The total noise temperature of the haloscope T$_{syst}$ from Eq.\eqref{eq:snr}, i.e. of the cavity and first amplifier stage, is an important parameter as the time required to obtain a given signal to noise ratio is proportional to T$_{syst}^2$. The follow up detectors of our path-finding haloscope described below will work at He$^3$/He$^4$ dilution temperatures. 
As for amplifiers, one of their main figures of merit is their input noise temperature, which contributes to the systems total noise. Josephson junction based amplifiers, such as Josephson parametric amplifiers (JPAs), have become the technology of choice for amplification near the quantum noise limit. GrAHal will benefit from the latest quantum amplifiers developments at Institut N\'eel: broad-band josephson travelling wave parametric amplifiers can be routinely fabricated using standard lithography techniques \cite{PhysRevX.10.021021}.



\subsection{Reachable axion-diphoton couplings}

\begin{figure}[h]
    \centering
    \includegraphics[width=0.48\textwidth]{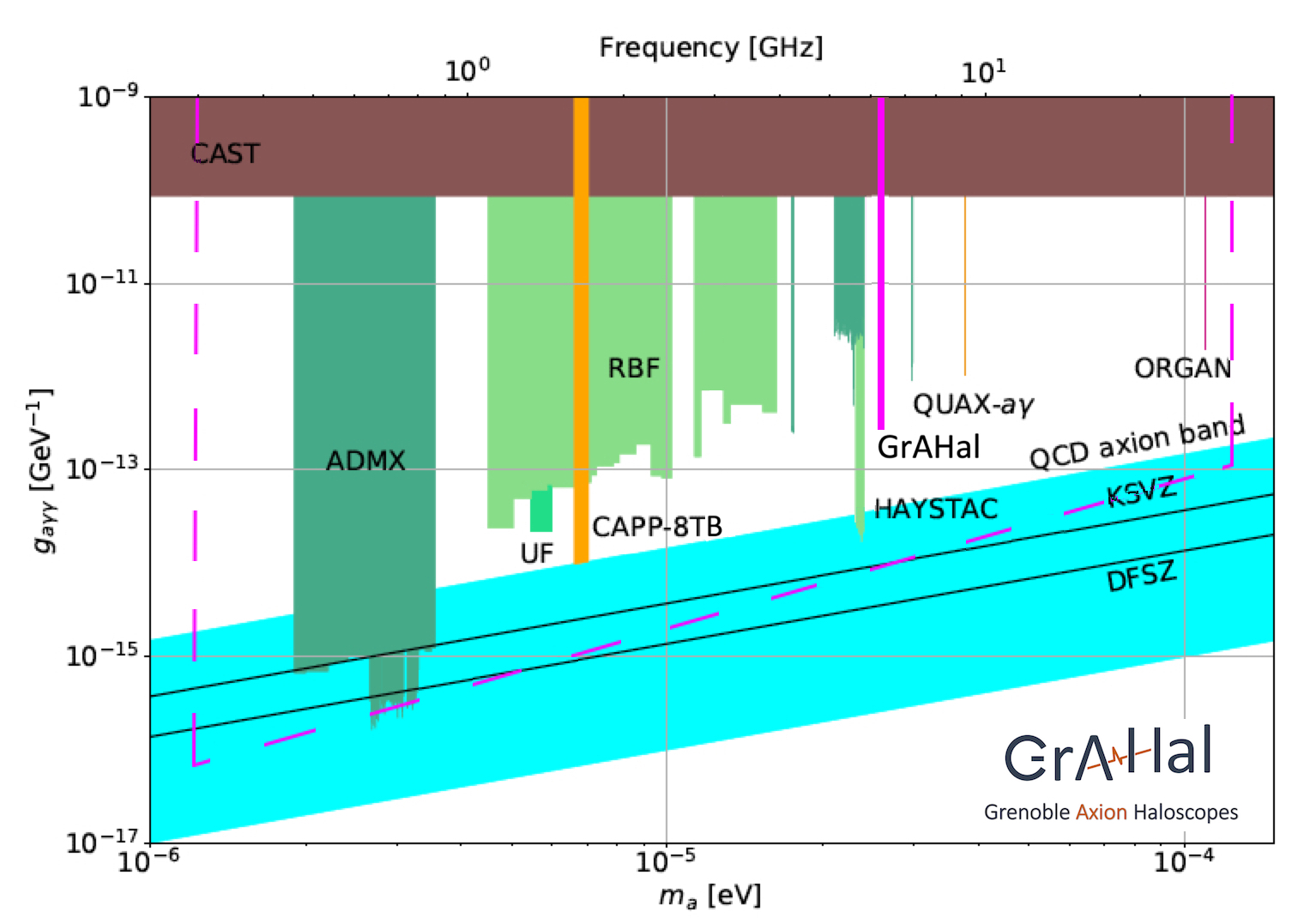}
     \caption{Exclusion limits for the axion diphoton coupling constant versus axion mass, presently established by haloscopes \cite{10.1093/ptep/ptaa104}. The blue band represents the area of theoretical interest based on the KSVZ and DFSZ models. The purple dashed line schematizes the exclusion domain which can be accessed by the GrAHal experiments, using the various magnetic field and RF cavity configurations described in the text. The vertical purple line represents the first exclusion limit obtained by the GrAHal path finding run at liquid He around 6.375 GHz (26.37 $\mu eV$). It will be improved by lowering the temperature and using a JPA.}
    \label{figure:exclusion diagram}
\end{figure}

As schematized in Fig.\ref{figure:exclusion diagram} a rather large axion-diphoton coupling exclusion domain will be accessible to the GrAHal experiments. However its full exploration with present technologies would involve a considerable integration time. GrAHal will first focus on essentially unexplored mass regions, namely 1.4-2 $\mu eV$ and 3.5-5 $\mu eV$ down to the DFSZ limit for the axion-diphoton coupling, as well as above 26 $\mu eV$ near the KSVZ limit. Further progress in detector sensitivity, combined with an increase of the number of haloscopes in operation and a worldwide coordination of experimental efforts will be of prime importance to reduce integration times.

\section{The pathfinding run}

GrAHal’s path-finding run aimed at designing and operating a first haloscope prototype working at liquid He temperature.
It consists of an oxygen free copper cylindrical cavity (inner diameter 36 mm, length 150 mm), presently without mechanical tuning system, with a TM$_{010}$ mode frequency close to 6.375 GHz at low temperature (axion mass close to $26.37 \mu \text{eV}$). It is maintained at liquid He temperature inside a 52 mm bore 14 Tesla magnet. Its two antennas (main and secondary ports) allow transmission and reflection characterization measurements. Its unloaded quality factor $Q_0$ is $37000 \pm 10\%$, the main antenna coupling constant is $\beta$ = 1.2 and we take $C$ = 0.69 for the TM010 mode coupling factor. The signal emerging from the cavity main port is first amplified by a Low Noise Factory cryogenic high-electron-mobility transistor amplifier, before further amplification and injection in a Rhode \& Schwarz ZVL13 analyzer for signal processing. The total gain of the receiver chain is calibrated using the power emitted by a heated load connected in place of the cavity. A few tuning steps could be taken by increasing the He exchange gas pressure. By increasing the inner dielectric constant of the cavity this allows an easily controlled downward tuning of the mode frequency, in a limited range.  
    \label{figure:GrAHal1_scheme}
The proper operation of the haloscope is checked by measuring fake axion signals injected in the cavity through the weakly coupled antenna port.

\begin{figure}
    \centering
    \includegraphics[width=0.4\textwidth]{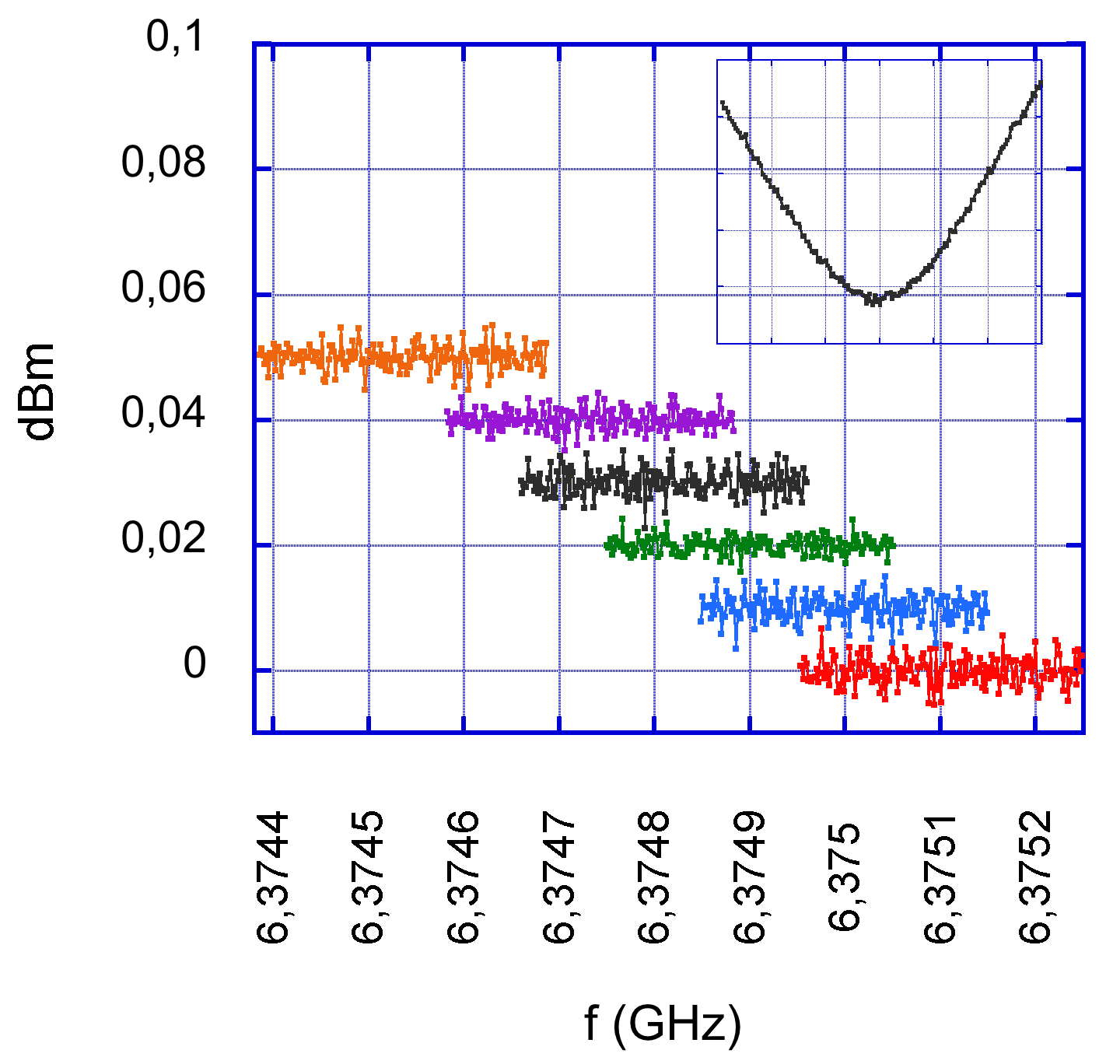}
    \caption{\underline{Main:} power spectra of the different tuning steps after subtraction of the background. Curves are shifted vertically for clarity. \underline{Inset:} raw spectrum with its background shape created by a small thermal imbalance between the cavity and the slightly warmer low T stage RF circuit.}
    \label{figure:raw-data}
\end{figure}

\begin{figure}
    \centering
    \includegraphics[width=0.45\textwidth]{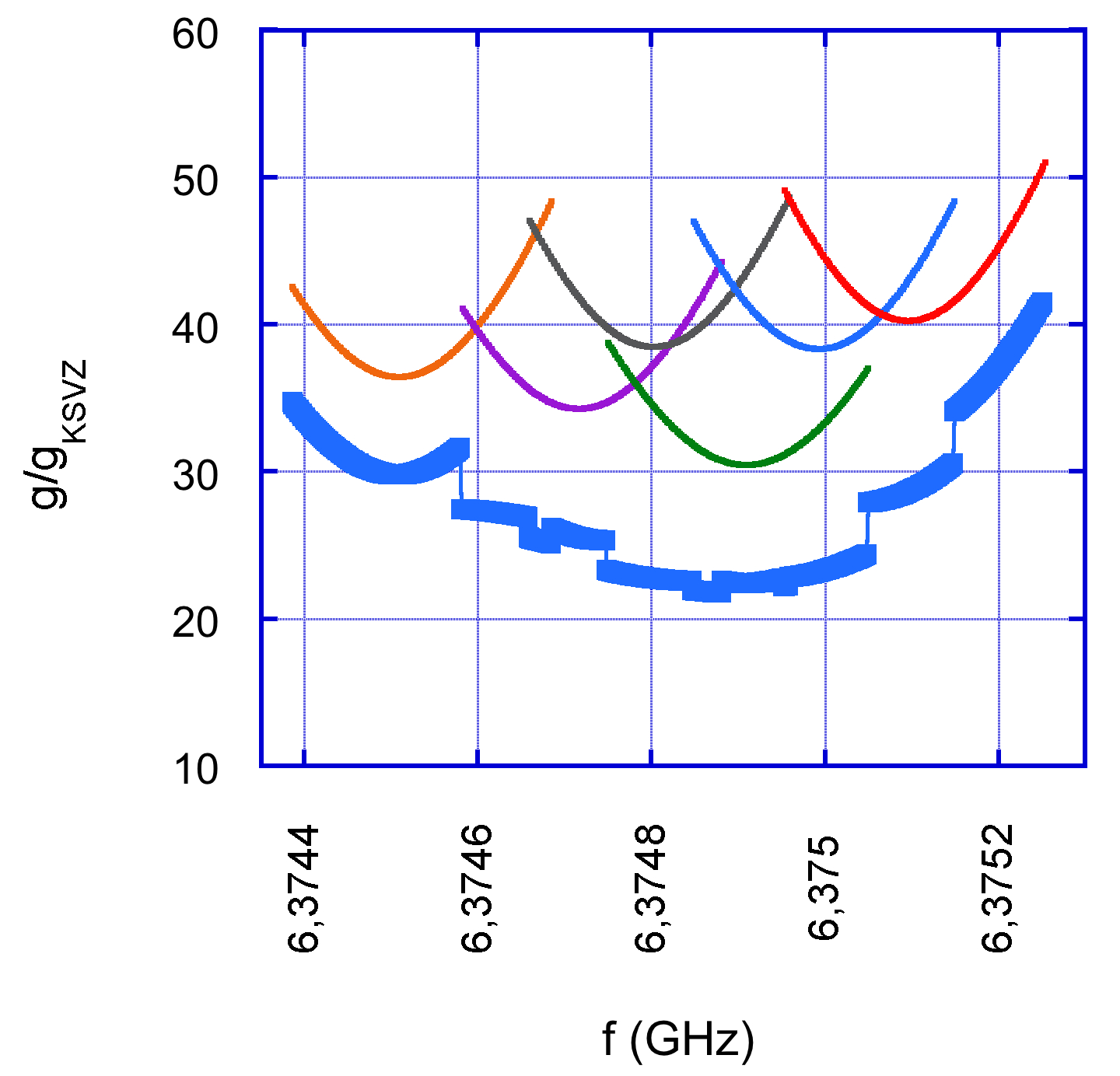}
    \caption{Individual (thin lines) and combined (thick blue line) exclusion limits on the axion-diphoton coupling relative to the KSVZ benchmark.}
    \label{figure:exclusion-limit}
\end{figure}

The path finding run took place during 6 days in July 2021.
Figure \ref{figure:raw-data} shows the data obtained after subtraction of the backgrounds.
We obtained a single combined spectrum by computing a weighted sum of the six flat spectra, taking into account their different integration times and frequency dependent detection sensitivity given by the lorentzian power resonance of the cavity in Eq.\ref{eq:power} \cite{PhysRevD.96.123008}, and applied a three points smoothing. For the axion peak detection, a $3\sigma$ threshold was applied to the combined spectrum. Any single point found above this value would constitute a rescan candidate. We performed numerical simulations adding synthetic virialized axion peaks of various heights to the data, and found that peaks of height $3.65 \sigma$ are detected at $95\%$ efficiency (i.e. $5\%$ of them produce no points above the $3\sigma$ threshold and are thus undetected). As no points were observed above the threshold, an exclusion limit was established for the axion-diphoton coupling constant, assuming axions constitute all of the local dark matter density. This limit is shown in Figure \ref{figure:exclusion-limit}.

\section{Conclusion}

The GrAHal collaboration has undertaken its first pathfinding run at liquid He temperature, which constrains axion-diphoton coupling to $g_{a \gamma \gamma} \leq 2.2\times10^{-13} \text{GeV}^{-1}$ ($g_{a \gamma \gamma} \leq 22 \times g_{KSVZ}$) around 6.375 GHz (axion mass of $26.37 ~ \mu\text{eV}$) with $95\%$ confidence. The subsequent developments of the project were briefly described and are underway. The GrAHal main goal is to develop a platform open to axion searches probing dark matter axions in the highly promising region of 0.3 - 30 GHz (1.25 - 125 $\mu\text{eV}$).

\subsection*{Acknowledgments} 

The authors would like to thank F. Caspers, A. Siemko, W. Venturini-Delsolaro and W. Wuensch from CERN for fruitful discussions,
C. Simon, Director of LNCMI, for his support and constructive approach toward sharing the future use of the Grenoble hybrid magnet between GrAHal and other scientific projects, David J. E. Marsh and the Grenoble ``Amis des Axions'' working group for helpful discussions. The pathfinder run was partially funded by the IDEX Universit\'e Grenoble Alpes and by the CNRS Institut National de Physique. The work of K.M., J.Q. and C.S. is supported by the IN2P3 Master project ``Axions from Particle Physics to Cosmology'' and the Enigmass Labex. J.Q.'s work is also supported by the IN2P3 Master project UCMN. The Grenoble hybrid magnet project is supported by Université Grenoble-Alpes (UGA), CNRS, French Ministry of Higher Education and Research in the framework of “Investissements pour l’avenir” Equipex LaSUP (Large Superconducting User Platform), European Funds for Regional Development (FEDER) and Rhône-Alpes region.

\addcontentsline{toc}{section}{References}
\bibliographystyle{JHEP}
\bibliography{refs}

\end{document}